# EXPERIMENTAL INVESTIGATION OF SPINNING MASSIVE BODY INFLUENCE ON FINE STRUCTURE OF DISTRIBUTION FUNCTIONS OF α-DECAY RATE FLUCTUATIONS.

V.A. Panchelyuga, S.E. Shnoll

Lomonosov's Moscow State University, Moscow, Russia
Institute of Theoretical and Experimental Biophysics RAS, Pushchino, Russia

*panvic333@yahoo.com, snoll@iteb.ru*

Abstract

The present investigation is dedicated to study of physical basis of macroscopic fluctuations effect [1]. In particular experimental investigation of possible influence of rapidly spinning massive body on distribution function of the α-decay rate fluctuations was carried out. Possible anisotropy of such influence was tested. The paper also contains fundamentals of the macroscopic fluctuations effect, method of experimental data processing and short review of phenomenology collected during more than fifty-years history of the macroscopic fluctuation effect investigations.

## Fundamentals of macroscopic fluctuations effect. Method of experimental data processing.

To understand the essence of macroscopic fluctuation effect let us consider a simple example. Suppose an electrical direct-current circuit. Also suppose that we provide a set of consecutive measurements of the current value, every time with more and more sensitive device. Then, at some point during such measurements we will be able to see that measured value (which was a constant at the beginning) is subjected to some fluctuations. Apparently obtaining fluctuations is possible this way in practically any process. Time series of fluctuations obtained in different processes are basic raw data for investigation of macroscopic fluctuations effect. Below we consider a method of experimental data processing, which is the basis for further investigations of macroscopic fluctuations effect.

This method can be divided into two stages. The first one is illustrated at the Fig. 1. Here Fig. 1A presents initial time series of fluctuations of some process. These initial time series are divided onto short intervals ordinarily of 30-60 points in length, Fig. 1B. For every such interval a histogram (distribution function of fluctuating values) is calculated, Fig. 1C. After this we smooth every histogram by *n*-points rectangular windows, Fig. 1D. The most often value of *n* is *n* = 4. As a result of the first stage of procedures applied to initial time series,

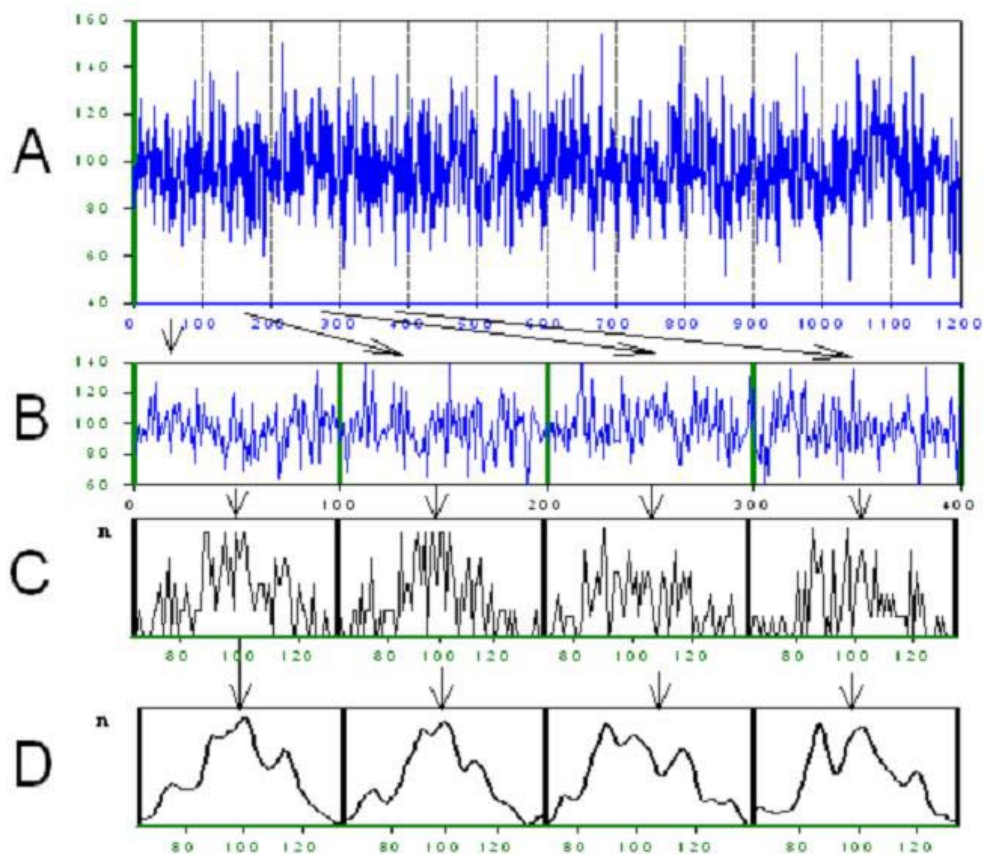

Fig. 1. Processing of initial time series. The method of smoothed histograms set obtaining.

Fig. 1A, we obtain a set of smoothed histograms, Fig. 1D. These histograms are subject of subsequent data processing procedure illustrated by Fig. 2 where the result of the second stage of data processing is presented.

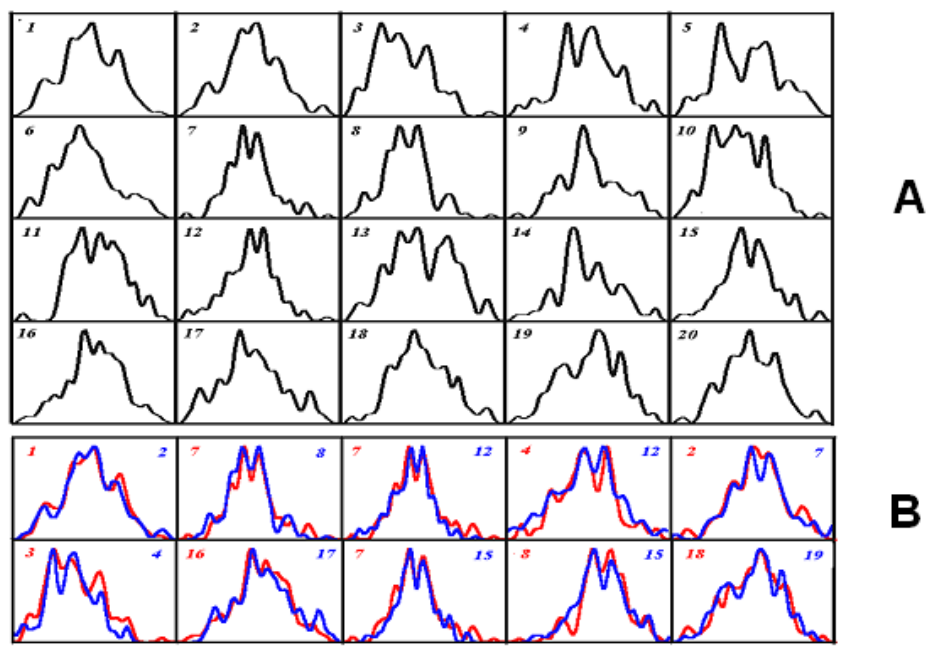

Fig. 2. Initial set of smoothed histograms and pairs of similar histograms.

Fig. 2A gives an example of $N = 20$ histograms set, which serves as initial material for the process of visual comparison of histogram pairs by an expert. The set at Fig. 2A, is obtained in the same way as the set at Fig. 1D. Every histogram of the set at Fig. 2A is compared with all other histograms of this set, or some other. In case when we compare histograms in the same set we obtain $N(N-1)/2$ comparisons of histogram pairs. In case when we compare histograms between different sets we obtain $N^2$ comparisons. For set displayed in Fig. 2A we obtained 190 comparisons. Fig. 2B presents 10 pairs histograms, which were found similar by expert.

It is possible to see from Fig. 2B that the process of histogram comparison by expert consists of evaluation of similarity of shape for histogram pairs. The process of expert histogram comparison is very sensitive to peculiarities of histograms shape. Usually the results of expert comparisons cannot be repeated by traditional methods of correlation analysis, spectral analysis, or using different measures of similarity, etc. [2]. Multiple attempts to create algorithm for automatic comparisons of histograms made clear that complete or partial automation of the process of expert histogram comparison is possible only using complex algorithms simulating some aspects of human perception, especially its whole nature.

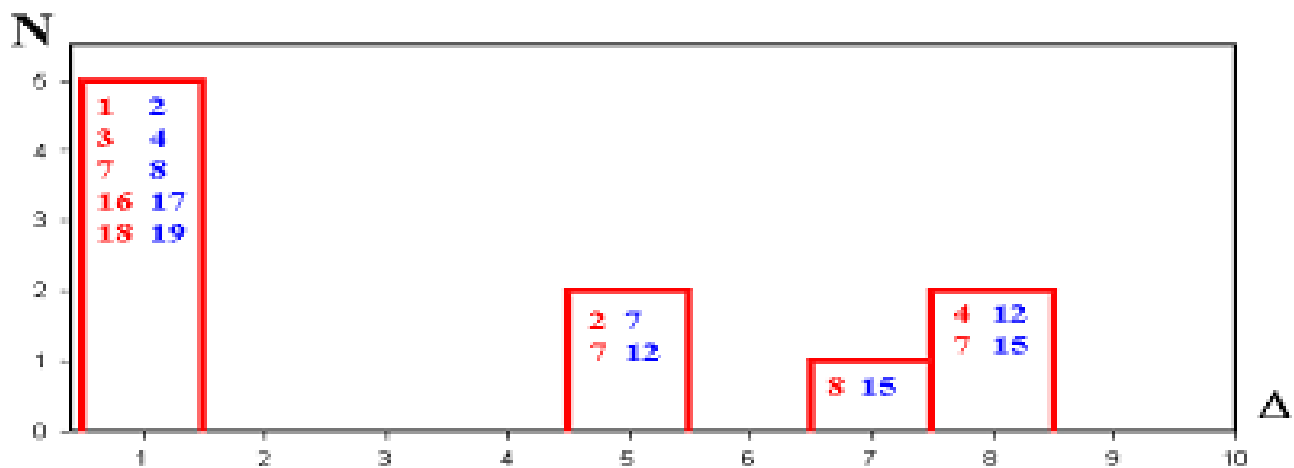

Fig. 3. An example of distribution of similar histograms by intervals between them for the set of histograms presented on the fig. 2A.

The end of the second stage is construction of distribution of number of similar histograms by time intervals between them. An example of such distribution for the set of histograms at Fig. 2A is presented at Fig. 3. Interval $\Delta$ is duration of time between each two histograms in time series. Expert estimation of similarity of histograms in pair is "1" or "0" value ("yes" or "no"). If histograms are similar, then similarity is equal to "1", in opposite case similarity is equal to "0". For example, the set of $N = 20$ histograms presented in Fig. 2A have $N - \Delta = 15$ pairs of histograms separated by interval $\Delta = 5$. From all of them only pairs №2-№7 and №7-

№12 are similar by expert's opinion. Consequently in the distribution, based on the set of histograms presented in Fig. 2A, number of similar histograms for interval $\Delta = 5$ will be equal N = 2.

Construction of interval distribution graph completes processing of experimental data by expert. On the basis of this distribution all main properties of macroscopic fluctuations effect are obtained. Following chapter gives a short revue of phenomenology of the effect.

Basic phenomenology of Macroscopic Fluctuations Effect.

The most general result of many-years investigations of macroscopic fluctuations effect is a proof of non-randomness of fine structure of histograms shapes built on the base of short samples of time series of fluctuations of different processes of any nature – from biochemical reactions and noises in gravitational antenna to fluctuations in $\alpha$-decay rate. Below we consider basic phenomenology of macroscopic fluctuations effect.

1. The Near Zone Effect.

The Near Zone Effect consists in higher probability of meeting similar pair of histograms in the nearest (neighboring) non-overlapping segments of time series of the results of measurements, Fig. 4 *a*). The effect leads to the notion of 'life-time' of histogram's definite shape. But at the present day it is not possible to point out time interval during which the shape of histogram is still invariable. The Near Zone Effect was tested for time intervals from several hours to seconds. Physical meaning of such a fractality needs further investigations. [1-3]

2. Universal Nature of Macroscopic Fluctuations Effect.

Universal nature of Macroscopic Fluctuations Effect means that the effect is invariant in relation to the qualitative nature of the fluctuation process. The facts of similarity of fine structure of histograms' shape in processes with energies differing in many orders (for example, energy of α-decay rate fluctuations and energy of noise in gravitational antenna differ approximately in 40 orders) mean that physical nature of this similarity is non-energetic. All above mentioned also represents a quite common reason of histograms' similarity. [1, 2, 4]

3. Periodical Manifestations of Macroscopic Fluctuations Effect.

Important evidence of non-randomness of histograms' shape is a regular character of its changes with time. This regularity manifests itself in the following phenomena.

3.1. Near-daily periods of changes in histograms shape similarity. They consist of two well-resolvable sidereal / star (1436 min) and solar (1440 min) periods. Existence of the periods means dependence of histograms shape on the rotations of the Earth around its axis.

3.2. Near-27-days periods of changes in histograms shape similarity. The periods probably mean dependence of histograms shape on the relative position of Earth, Moon, Sun and, probably, the planets. [7]

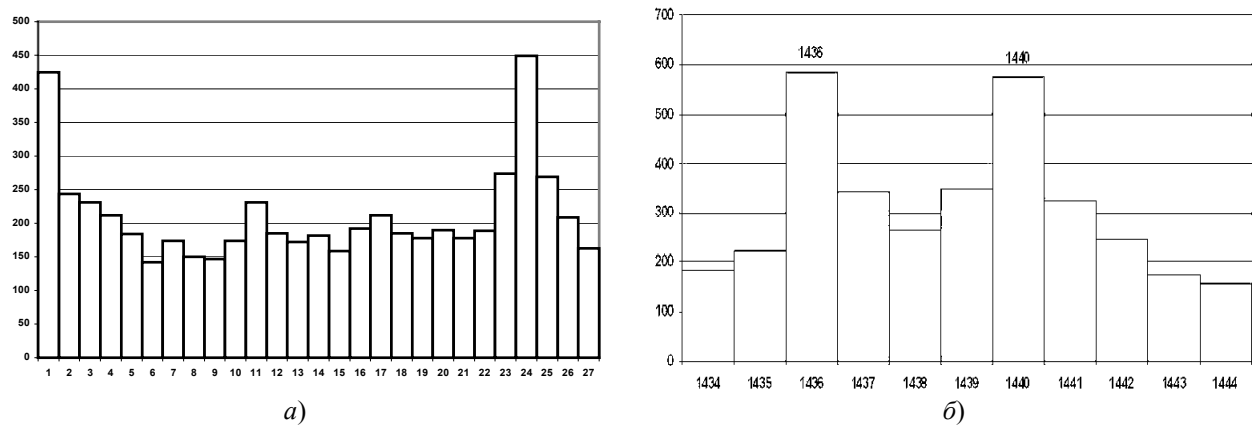

Fig. 4. *a*) Example of Near Zone Effect and daily period [6], *b*) splitting of daily period on solar and star periods [7]. X-axis – time intervals between pairs of similar histograms, *a*) – hours, *b*) – minutes, Y-axis – number of similar histogram pairs found by expert.

3.3. Yearly periods of changes in histograms shape similarity. They consist of solar (365 solar days) and sidereal (365 solar days, 6 hours and 9 minutes) periods. [8]

All above-mentioned periods mean dependence of histograms shape on the rotations of the Earth around its axis and movements of the Earth along its circumsolar orbit.

4. Local-Time Effect.

Dependence of histograms' shape on the rotations of the Earth around its axis manifests itself in the local-time effect. The effect consists in synchronous changes of histograms' shape similarity for different geographical locations at the same local time. It was tested many times for different geographical locations around the Globe. It was found that the effect works for maximally possible distances (about 15000 km) between the places of measurements. For some cases absolute-time synchronism (synchronous changes of histograms' shape similarity for different geographical locations at the same moments) can be observable.

Fig. 5 presents two intervals distributions constructed on the base of time series of α-decay rate fluctuations of $^{239}$Pu. The time series were obtained in Moscow region, (Pushchino, latitude 54°50′ North and longitude 37°38′ East) and in Antarctica (Novolazarevskaja station, latitude 70°02′ South and longitude 11°35′ West). The distance between the points of measurements is about 14500 km and difference in local time is 103 min.

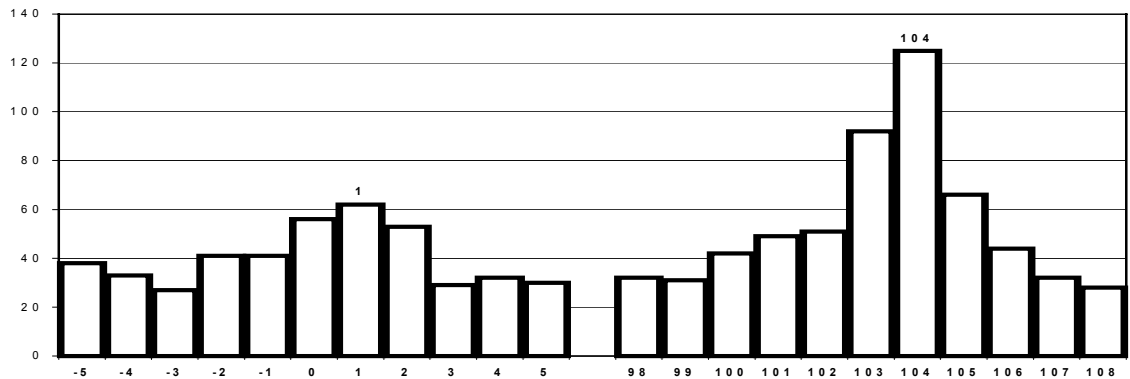

Fig. 5. Synchronous changes of histograms' shape similarity in different geographical locations. Left intervals distribution presents the effect of histogram similarity by absolute time, right distribution presents similarity by local time.

The left side of fig. 5 presents interval distribution, which illustrates the effect of histogram similarity by absolute time. The right side of fig. 5 presents the effect of similarity by local time. It can be seen that local time effect appears more clearly. [1, 2]

5. Disappearance of daily periods for measurements near the North Pole.
The dependence of histograms' shape on the Earth rotation around its axis is also revealed in disappearance of daily periods in measurements conducted close to the North Pole. Such measurements were carried out at the latitude 82° North in 2000. Near-daily periods disappeared for histograms in 15-minut and 60-minut length. But for 1-minute histograms the periods were found. For such histograms a local-time effect was also found. [7]

Above-mentioned results lead to necessity of measurements as close as possible to the North Pole. Impossibility of such measurements stimulates us to start measurements with collimators cutting out a stream of α-particles at radioactive decay of $^{239}$Pu. Results of these experiments radically change our understanding of macroscopic fluctuations effect. [9]

6. Motionless collimator directed at the Polar Star.
Measurements were taken with the collimator of α-particles directed at the Polar Star. For these measurements the near-daily periods and near zone effect was not observed. The measurements were made in Pushchino at latitude 54° North, but the effect was as would be expected at latitude 90° North, i.e. at the North Pole. This indicates the dependence of histograms shape on the direction in space. Such a dependence in its turn leads to the conclusion about anisotropy of space itself. [7, 9]

## 7. Motionless collimators directed to the East and to the West.

The conclusion about anisotropy of space was confirmed by measurements with two collimators. One of them was directed to the East; the other one to the West. In those experiments two important effects were discovered.

7.1. The histograms registered in the experiments with the East-directed collimator are similar to histograms from West-directed collimator with delay of 718 min, i.e. half of the sidereal day.

7.2. No similar histograms were observed in the simultaneous measurements with the East and West collimators. Without collimators, it is highly probable for similar histograms to appear at the same place and time. This space-time synchronism disappears when α-particles streaming in the opposite directions are counted.

These results are in agreement with the concept that the histogram shape depends on the direction of the α-particle emission i.e. with the concept of space anisotropy. [10]

## 8. Experiments with the rotating collimators.

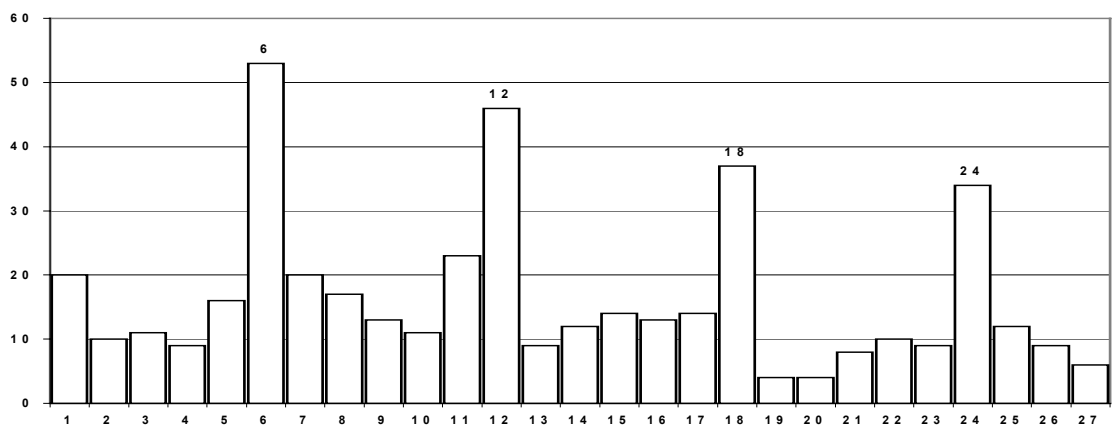

Fig. 6. Interval distribution obtained on the base of 60-min histograms constructed from measurements of α-decay rate fluctuations of collimated $^{239}$Pu source.

Experiments with rotating collimators were a natural development of above-mentioned investigations. [11]

8.1. Collimator rotating counter-clockwise scans coelosphere with the period equal number of collimator rotations plus one rotation made by Earth itself. The dependence of the probability of appearance similar histograms on the number of collimator rotations per day was studied. Just as expected, the probability jumps with periods equal to 1440 min divided by the number of collimator rotations per day plus 1. Examination of experimental data at 1, 2, 3, 4, 5, 6, 7, 11 and 23 rotations per day reveals periods equal to 12, 8, 6 etc. hours. The analysis of 1-min histograms shows that each of these periods has two extremes: "sidereal" and "solar". These results indicate that the histogram pattern is indeed determined by direction of α-particle emission. [11] An example of 6-hours period obtained with 60-min histograms at three counter-clockwise rotations of collimator per day is presented in fig. 6

8.2 For collimator, which made 1 clockwise rotation per day, the rotation of the Earth was compensated (α -particles always emitted in the direction of the same region of the coelosphere) and, correspondingly, the daily periods disappeared. This result was completely analogous to the results of measurements near the North Pole and measurements with the immobile collimator directed towards the Polar Star [10].

8.3 With the collimator placed at the ecliptic plane, directed toward the Sun and making 1 clockwise rotation per day, α-particles are constantly emitted in the direction of the Sun. As it was expected, the near-daily periods, both solar and sidereal, disappeared in such conditions.

9. Characteristic histograms' shapes at New Moon and solar eclipses.

All the results presented above have probabilistic character and were obtained by the evaluation of tens of thousands of histogram pairs in every experiment. A completely different approach is used in the search for characteristic histogram shapes in the periods of the New Moon and solar eclipses. In these cases the histograms' shape is examined at a certain predetermined moment of New Moon or solar eclipse. In such a way it was discovered that at the moment of the New Moon, a certain characteristic histogram appears practically simultaneously at different longitudes and latitudes — all over the Earth. This characteristic histogram corresponds to a time segment of 0.5–1.0 min [12]. When the solar eclipse reaches maximum (as a rule, this moment does not coincide with the time of the New Moon), a specific histogram also appears; however, it has a different shape. Such specific shapes emerge not only at the moments of the New Moon or solar eclipses, though the probability of their appearance at these very moments at different places and on different dates (months, years) is extremely low. These specific histograms' shapes neither relate to tidal effects nor depend on position on the Earth's surface, where the Moon's shadow falls during the eclipse or the New Moon.

10. Characteristic histograms shapes at rise and set of Sun and Moon.

The shape of histograms is determined by a complex set of cosmo-physical factors. As it follows from the existence of the near-27-day periods, amongst these factors may be the relative positions and states of the Sun, the Moon and the Earth. We repeatedly observed similar histograms during the risings and settings of the Sun and the Moon. A very large volume of work has been carried out. Yet we have not found a histogram shape, which would be characteristic for those instants. A review and analysis of the corresponding results will be given in a special paper.

11. Mirror symmetry of histograms.

Very often (up to 30%) shape of histograms in the similar pair has "mirror" symmetry. This means an existence of left and right shapes. This phenomenon possibly signifies that chirality is an immanence property of space-time. [1]

The possible nature of the Macroscopic Fluctuations Effect. Idea of the present investigation.

Above-mentioned properties №3 - №4 of the macroscopic fluctuations effect pointed out the dependence of the effect phenomenology on the space position of the Earth, Moon, and the Sun and property №2, which states independence of the phenomenology from qualitative nature of fluctuating process, lead to supposition that phenomenology can be determined by only a such common factor as space-time heterogeneity. The space-time heterogeneity can be connected with gravitational interaction. On the other hand, properties №5 - №8 indicating space anisotropy of acting agent and properties №9 - №10 – synchronous arising of similar histograms shapes in different geographical locations at certain moments in dynamics of Sun, Earth, and Moon leads to conclusion about wave nature of acting agent.

To sum up we can suppose that acting agent determining above-mentioned properties of macroscopic fluctuations have gravity-wave nature. According to this, shape of histograms, can be sensitive to gravitational wave influence. This is the base idea of experiment, which is presented at fig. 7.

Experimental setup.

At fig. 7 simplified diagram of experimental setup on detecting gravitational wave influence on the shape of α-decay rate histograms is presented. The left side of the diagram schematically presents generator of gravitational influence. A centrifuge K70 symmetrically loaded with two bottles of water was used as such generator. The weight of every bottle was 1.5 kg

Gravitational radiation of the generator, schematically presented by parallel arrows, influences on two-channel registration system, showed as Ch. 1 and Ch. 2. The system consists of two recorders of α-decay rate from $^{239}$Pu-sources. Average α-decay rate for Ch. 1 is 272 decays per second and 174 decays per second for Ch. 2. The recorders lie in the plane of centrifuge rotor and are placed at a distance of 1.5 m from its axle. For every recorder the angle φ between wave vector of generating gravitational wave and direction of α-particles emitting is different. For Ch. 1-recorder the angle is φ = 180° and for Ch. 2 - φ = 90°. In view of wave nature of expected influence it must be angle-sensitive. So, recorders Ch. 1 and Ch. 2 must be of different sensitivity to generate wave influence. Values of α-decay rate per second from every channel and speed of rotation of centrifuge rotor were registered by a special computer system.

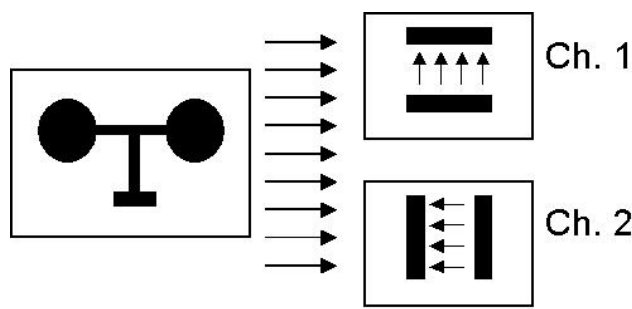

Fig. 7. Simplified diagram of experimental setup on detecting gravitational wave influence on the shape of α-decay rate histograms

Experiments were carried out as 5-minute cycles of running and turning off the centrifuge. So, period of influence was equal to 10 min. The rotation speed of running centrifuge rotor was 3000 revolutions per minute. The rotor of turning off centrifuge at the end of 5-minute cycle keeps the rotation speed about 300 rpm.

Simulation of expected results.

Fig. 8 *a*) shows idealized diagram presenting change of rotation speed of centrifuge rotor with time. We expect, that all histograms constructed from pieces of time series corresponding to running centrifuge are similar between themselves, but non-similar to histograms constructed from pieces of time series corresponding to turning off centrifuge. In the same way all histograms constructed from pieces of time series corresponding to turning off centrifuge are similar between themselves, but non-similar to histograms constructed from pieces of time series corresponding to running centrifuge.

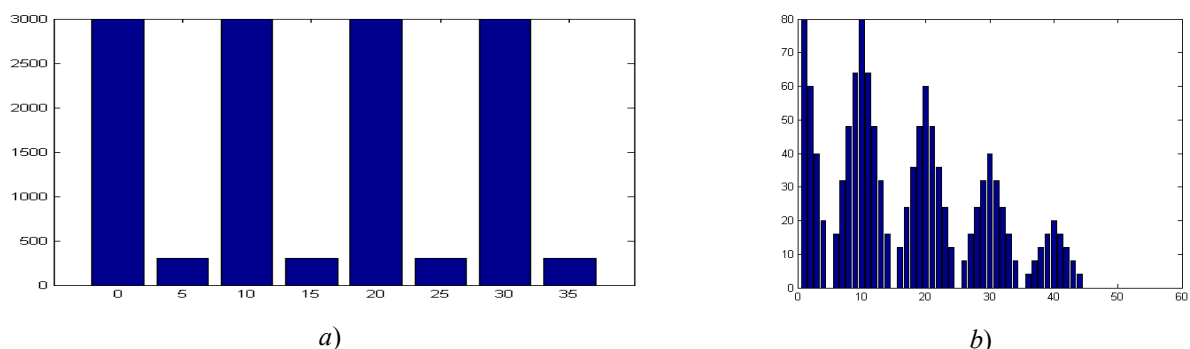

*a)*          *b)*

Fig. 8. Idealized diagram of rotation speed of centrifuge rotor with time, *a*); expected interval distribution, *b*).

Above supposition allows us to calculate expected interval distribution, fig. 8 *b*). As it can be seen at fig. 8 *b*) interval distribution for experimental record of fixed length consists of finite number of decreasing peaks repeating with period, which equal to period of centrifuge alternating.

Experimental results.

According to above described method five series of measurements were carried out. An example of experimental record No. 4 obtained from Ch. 1-recorder is given at fig. 9 *a*). This graph presents a piece in 2500 sec length of time series in 26400 one-second measurements. Fig. 9 *b*) presents distribution function for this time series. As it is possible to see from fig. 9 *a*) and fig. 9 *b*) time series of $\alpha$-decay rate fluctuations and its distribution function are typical for this process. Absence of any peculiarities in the presented time series and the distribution function are expected and is evidence of good quality of experimental registration. As it was noted at the beginning, traditional methods of time series processing are not sensitive to macroscopic fluctuations effect manifestations.

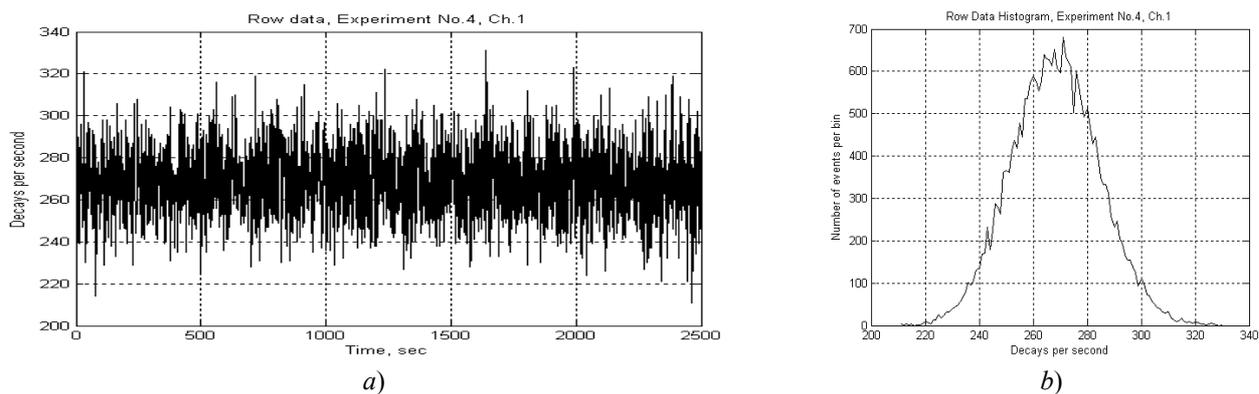

Fig. 9. An example of experimental record of $\alpha$-decay rate fluctuations obtained from $^{239}$Pu-source (Ch. 1, series No. 4), *a*); and corresponding distribution function, *b*).

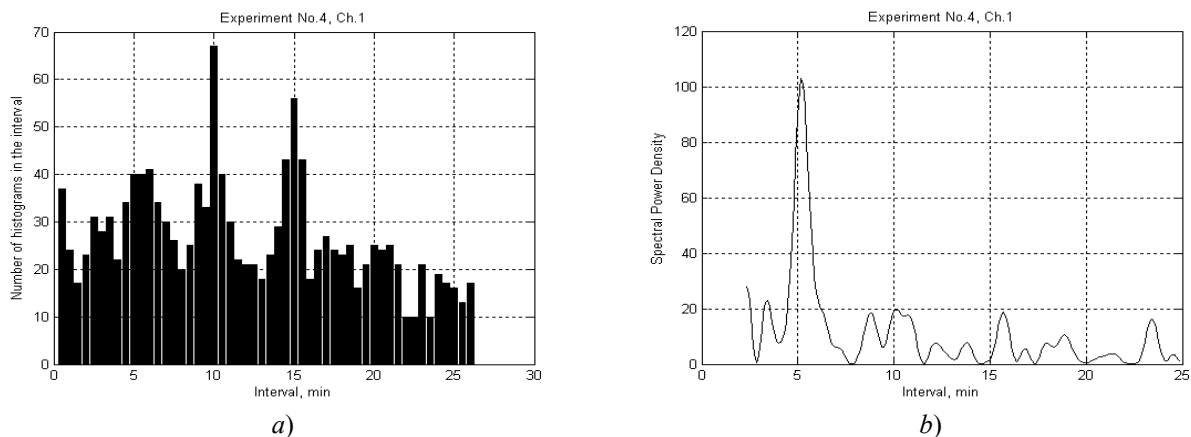

Fig. 10. An example of interval distribution (Ch. 1, series No. 4), *a*); and corresponding density function of power spectrum, *b*).

According to the methods described in first chapter, on the base of obtained experimental records five sets of 0.5-min histograms were constructed. Expert tested the pairs of histograms from every set for similarity. Typical example of interval distribution constructed on the base of result of expert comparison (Ch. 1, series No. 4) is presented at fig. 10 *a*).

As it is possible to see from fig. 10 the interval distribution consists of quite distinct periodic peaks. The period of peak repetition is 5 min. Fig. 10 *b*) presents spectral power density corresponding to the interval distribution. As it can be expected from interval distribution the spectrum has distinct 5-minute peak. Appearance of the 5-minute period is quite unexpected from the point of view of above developed model. The meaning of the period will be considered below.

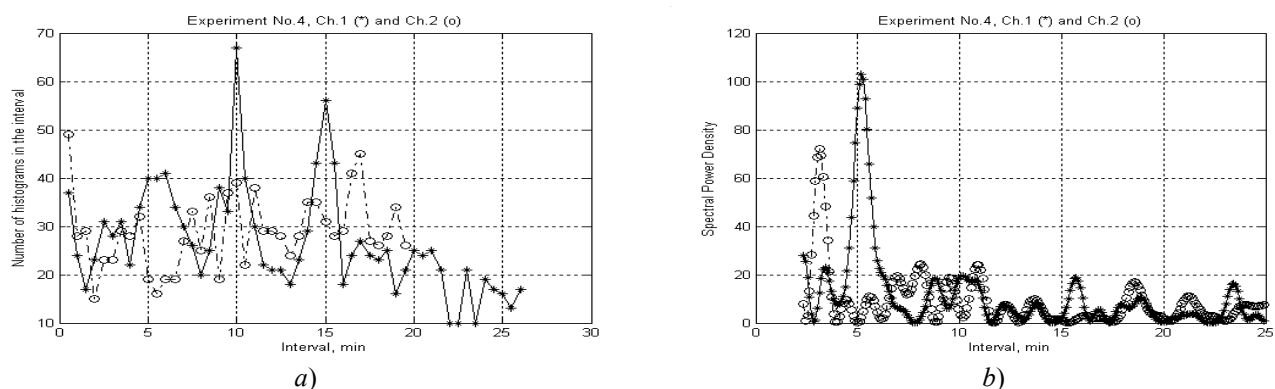

Fig. 11. An Example of interval distribution for Ch. 1(*) and Ch. 2 (°) (series No. 4), *a*); and corresponding densities of power spectrum, *b*).

For convenience in fig. 11 *a*) interval distributions for Ch. 1 (marked by asterisks) and Ch. 2 (marked by little circles) are given. As it is possible to see, periodical pattern typical for Ch. 1 is absent in Ch. 2. Fig. 11 *b*) presents spectral power densities corresponding to interval distributions in fig. 11 *a*). It is clear that 5-minute peak is absent for Ch.2 spectrum. This result validates supposition that registration system is angle-sensitive in relation to generated influence. At the same time for some cases spectrum for Ch. 2 contains 2.5-3 - minute peak, which can also be seen in fig. 11 *b*). The physical nature of this peak and its correspondence to centrifuge dynamic is unknown.

Acceleration modes.

Results, illustrated above by data for series No. 4 were also obtained for other series. This allows us to make a statement about sensitivity of histograms' shape to influence of rotating centrifuge rotor. This influence reveals itself by higher probability of meeting similar pair of histograms with period, which equals to 5 minute.

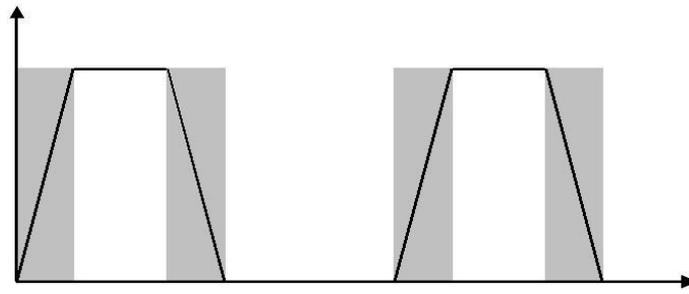
Fig. 12. Simplified diagram for centrifuge rotor acceleration and braking.

Appearance of 5-minute period instead of 10-minute one indicates that histograms shape is sensitive not to rotation speed of centrifuge rotor, as it was suggested in the model presented at fig. 8, but to its accelerations. Fig. 12 illustrates this supposition. Here gray rectangles mark intervals of acceleration and braking of centrifuge rotor. Every period includes two such intervals. If histograms shape is sensitive to accelerations, we will obtain double frequency, i.e. 5-minute period instead of 10-minute one. Interval distribution in this case will be the same as presented in fig. 8 *b*) but with 5-minute peaks period, which is observed in interval distribution obtained from experimental data.

Experiments with rotation speed of centrifuge rotor confirm this supposition. The upper graph in fig. 13 presents test record of three acceleration / braking periods of rotation speed of centrifuge rotor. The lower graph in fig. 13 presents derivative of rotation speed, which corresponds to rotor accelerations. Narrow peaks in this graph correspond to acceleration mode. We suppose that histograms of α-decay fluctuations, which correspond to acceleration modes in centrifuge operations determine 5-minute period observable in above described experiments.

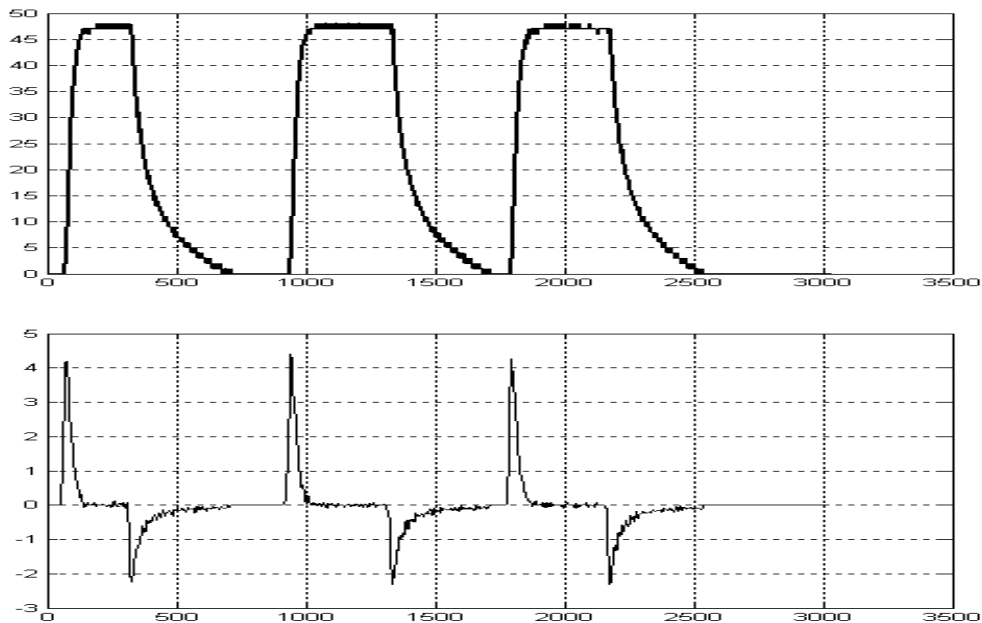
Fig. 13. Upper graph: tests record of three acceleration-braking periods of rotation speed of centrifuge rotor; lower graph: derivative of rotation speed (acceleration), presented at the upper graph.

Discussion.

In favor of the supposition that acceleration modes can define the shape of histograms we shall consider works [13, 14]. These works used a registration system providing acceleration modes, which are in some way complementary to such modes in our works. This registration system used as a sensor rapidly spinning massive body with artificially created acceleration mode by means of special braking pulse. Duration of the pulse is 18-30% of rotation period. [13] Registering parameter of the system is angular velocity of spinning body. It turned out, that such a system is sensitive to the same events, which are noted in properties No. 9 and No. 10 of macroscopic fluctuation effect phenomenology. [14] All this events have a relation to certain extrema in the velocity of change of the space-time position of the Sun, the Earth and the Moon; in this respect, the situation can be considered as a regime with acceleration and thus, it can determine the form of histograms describing the fluctuations in various processes.

As the second example of experimental investigation, where acceleration modes play an important role, we will consider work [15]. In this work a pair of generator with identical crystal oscillators was used as a registration system. The generators were placed in such a way that the position of the crystals were orthogonal. Registering parameter of this system is relative change of resonant frequencies of crystals of the generators. Authors named this parameter a T-signal. A study of daily changes of T-signal shows its anisotropy with extrema at local noon and midnight. Authors note non-electromagnetic nature of T-signal, and its biological activity. The T-signal changes were considered as consequence of gravitation waves emission of the Sun.

It is possible to note the general moments typical for works [13-15] and experiment considered in the present work. The first important feature is presence of "acceleration modes" in the registering system, allowing distinguishing some direction in space. In our experiments it is set by a direction of outgoing α-particles, in [13-14] – by the moment of breaking impulse, in [15] – by a perpendicular to a plane of the crystal plate oscillations.

"Acceleration modes", causing anisotropic properties of registering system at the same time make it sensitive to the same "acceleration modes" which are external in relation to it and, presumably, are connected with gravity-wave radiation.

Summarizing we shall note, that as a result of the present experimental investigation influence of quickly rotating massive body on the shape of the fine structure of constructed upon small samples distribution functions of fluctuations of α-decay rate, appearing in the higher probability of similarity of shape of the histograms for the moments corresponding to "acceleration modes" is fixed. The influence possesses anisotropic properties and, presumably, has the gravity-wave nature.

The authors are grateful to N.V. Udaltsova for valuable help in preparation of text of the paper.

REFERENCES.

1. S.E. Shnoll, V.A. Kolombet, E.V. Pozharskii, T.A. Zenchenko, I.M. Zvereva, A.A. Konradov "Realization of discrete states during fluctuations in macroscopic processes", Physics-Uspekhi 41(10), 1025-1035 (1998)
2. S.E. Shnoll, T.A. Zenchenko, K.I. Zenchenko, E.V. Pozharskii, V.A. Kolombet, A.A. Konradov "Regular variation of the fine structure of statistical distributions as a consequence of cosmophysical agents" Physics – Uspekhi 43(2) 205-209 (2000)
3. S.E. Shnoll, N.V. Udaltsova, V.A. Kolombet, V.A. Namiot, and N.B. Bodrova Regularities in the discrete distributions of the results of measurements (cosmophysical aspects). Biophysics, 1992, v. 37(3), 378–398.
4. S.E. Shnoll, V.A. Namiot, V.E. Zhvirblis, V.N. Morozov, A.V. Temnov, and T.Ya. Morozova Possible common nature of macroscopic fluctuations of the rates of biochemical and chemical reactions, electrophoretic mobility of the cells and fluctuations in measurements of radioactivity, optical activity and flicker noise. Biophysics, 1983, v. 28(1), 164–168.
5. S.E. Shnoll The form of the spectra of states realized in the course of macroscopic fluctuations depends on the rotation of the Earth about its axis. Biophysics, 1995, v. 40(4), 857–866.
6. S.E. Shnoll Discrete distribution patterns: arithmetic and cosmophysical origins of their macroscopic fluctuations. Biophysics, 2001, v. 46(5), 733–741.
7. S.E. Shnoll Periodical changes in the fine structure of statistic distributions in stochastic processes as a result of arithmetic and cosmophysical reasons. Time, Chaos, and Math. Problems, No. 3, University Publ. House, Moscow, 2004, 121–154.
8. Shnoll S.E., Zenchenko K.I., Udaltsova N.V., Cosmo-physical effects in structure of the daily and yearly periods of change in the shape of the histograms constructed by results of measurements of alpha-activity $Pu^{239}$. http://arxiv.org/abs/physics/0504092
9. S.E. Shnoll, K.I. Zenchenko, I.I. Berulis, N.V. Udaltsova, S.S. Zhirkov, and I.A. Rubinstein The dependence of "macroscopic fluctuations" on cosmophysical factors. Spatial anisotropy. Biophysics, 2004, v. 49(1), 129–139.
10. Simon E. Shnoll, Konstantin I. Zenchenko, Iosas I. Berulis, Natalia V. Udaltsova and Ilia A. Rubinstein Fine structure of histograms of alpha-activity measurements depends on direction of alpha particles flow and the Earth rotation: experiments with collimators. http://arxiv.org/abs/physics/0412007

11. Shnoll S.E., Rubinshtejn I.A., Zenchenko K.I., Shlekhtarev V.A., Kaminsky A.V., Konradov A.A., Udaltsova N.V. Experiments with Rotating Collimators Cutting out Pencil of α-Particles at Radioactive Decay of $^{239}$Pu Evidence Sharp Anisotropy of Space // Progress in Physics, V. 1, 2005, pp. 81-84. http://arxiv.org/abs/physics/0501004
12. Shnoll S.E., Zenchenko K.I., Shapovalov S.N., Gorshkov S.N., Makarevich A.V. and Troshichev O.A. The specific form of histograms presenting the distribution of data of α-decay measurements appears simultaneously in the moment of New Moon in different points from Arctic to Antarctic. http://arxiv.org/abs/physics/0412152
13. B.Yu.Bogdanovich, I.S. Shchedrin, V.N. Smirnov, N.V. Yegorov Special method of massive body spinning as instrument for astrophysical research. Scientific Session of MIFI, 2003, v.7, 45-46.
14. B.Yu.Bogdanovich, I.S. Shchedrin, V.N. Smirnov, N.V. Yegorov Preliminary estimations of kinetic energy changes of spinning massive body on space-time position of Sun and Moon. Scientific Session of MIFI, 2003, v.7, 47-48.
15. N.V. Klochek, L.E. Palamarchuk, M.V. Nikonova Preliminary results of investigation of the influence of cosmophysical radiation of non-electromagnetic nature on physical and biological systems. Biophysics, 1995, 40(4), 889-896.